\documentclass[aps,tightlines]{revtex4}

\usepackage{amsfonts}
\usepackage{amsmath}
\usepackage{amssymb}
\usepackage{graphicx}

\begin{document}

\title[Negativity Fonts...]{Negativity Fonts, multiqubit invariants and Four
qubit Maximally Entangled States}
\author{S. Shelly Sharma}
\email{shelly@uel.br}
\affiliation{Departamento de F\'{\i}sica, Universidade Estadual de Londrina, Londrina
86051-990, PR Brazil }
\author{N. K. Sharma}
\email{nsharma@uel.br}
\affiliation{Departamento de Matem\'{a}tica, Universidade Estadual de Londrina, Londrina
86051-990 PR, Brazil }
\thanks{}

\begin{abstract}
Recently, we introduced negativity fonts as the basic units of multipartite
entanglement in pure states. We show that the relation between global
negativity of partial transpose of N- qubit state and linear entropy of
reduced single qubit state yields an expression for global negativity in
terms of determinants of negativity fonts. Transformation equations for
determinants of negativity fonts under local unitaries (LU's) are useful to
construct LU invariants such as degree four and degree six invariants for
four qubit states. The difference of squared negativity and N-tangle is an N
qubit invariant which contains information on entanglement of the state
caused by quantum coherences that are not annihilated by removing a single
qubit. Four qubit invariants that detect the entanglement of specific parts
in a four qubit state are expressed in terms of three qubit subsystem
invariants. Numerical values of invariants bring out distinct features of
several four qubit states which have been proposed to be the maximally
entangled four qubit states.
\end{abstract}

\maketitle


\section{Introduction}

Entanglement is an intriguing property of quantum systems and its detection,
characterization and quantification are important questions in quantum
mechanics. For a pure state of bipartite quantum system consisting of two
distinguishable subsystems $A$ and $B$, each of arbitrary dimension,
negativity \cite{zycz98,vida02}, and linear entropy calculated from reduced
density operator of either element, may be chosen as entanglement measures.
For tripartite case, besides the quantity of entanglement we must also know
whether the entanglement is GHZ-like or W-like\ \cite{dur00} and states are
grouped into distinct entanglement classes for four qubits \cite%
{vers02,vers03,acin00,acin01,miya03,miya04}. An entanglement measure must
have value in the range zero for the product state to a maximum value for a
maximally entangled state and satisfy the minimal requirement of local
unitary invariance \cite{vida00}. Generally accepted measures of
entanglement, such as concurrence \cite{woot98} for two qubits, and three
tangle \cite{coff00} for three qubits, turn out to be such invariants \cite%
{vers03,ging02}. In the case of four qubits, the standard approach from
invariant theory, employing the well established W-process by Cayley, has
lead to the construction of a complete set of SL-invariants \cite{luqu03}.
In ref. \cite{luqu06} the invariants up to degree 6 have been determined
together with 5 invariants of degree 8. Local unitary invariants have been
reported \ for even number of qubits in ref. \cite{wong01} and for even and
odd number of qubits in \cite{li06,li07,li10}. Independent of these
approaches, a method based on expectation values of antilinear operators
with emphasis on permutation invariance of the global entanglement measure 
\cite{oste05,oste06}, has been suggested. Permutation invariance has been
highlighted as a demand on global entanglement measures already in Ref. \cite%
{coff00} and later in Ref. \cite{chte07}.

Negativity of global partial transpose is a widely used computable measure
of free bipartite entanglement. Negativity is based on Peres-Horodecki NPT
criterion \cite{pere96,horo96} and is known to be an entanglement monotone 
\cite{vida02}. A global partial transpose with respect to a sub system $p$
is obtained by transposing the state of subsystem $p$ in state operator.\ In
refs. \cite{shar101,shar102}, we introduced negativity fonts defined as two
by two matrices of probability amplitudes that determine the negative eigen
values of four by four submatrices of partially transposed state operators.
It was shown that relevant $N-$qubit local unitary invariants can be
obtained, directly, from transformation properties of determinants of
negativity fonts under local unitary transformations. From expression of an
invariant in terms of determinants of negativity fonts, one can easily read
how subsystems invariants contribute to the composite system invariant. In
this article, we obtain an expression for global negativity in terms of
determinants of negativity fonts. The squared negativity of $N-$qubit
partially transposed operator, is found to be the sum of squares of moduli
of determinants of all possible negativity fonts. For the sake of
completeness, we briefly outline the procedure for constructing two-qubit
local unitary (LU) invariants for an $N-$ qubit state by examining the
intrinsic sources of negativity present in global and $K-$way partially
transposed matrices. In a four qubit state, the entanglement of a three
qubit subsystem may arise due to four-way or three-way correlations. We show
that four qubit invariants that detect the entanglement of two qubits in a
four qubit state \cite{shar102} are combinations of three qubit invariants.
A form of degree six invariant for four qubit states constructed in terms of
negativity font determinants demonstrates the ease with which complex
invariants can be written down from basic principles and calculated
numerically. Numerical values of invariants are found to bring out distinct
features of several known four qubit states which have been proposed to be
the maximally entangled states.

Definition of negativity fonts and the notation to represent determinants of 
$N-$way and $K-$way negativity fonts is given in section II. Transformation
equations for determinants of negativity fonts are used to obtain an
expression for square of global negativity in terms of determinants of
negativity fonts in section III. Section IV details degree two, four and six
invariants for a generic four qubit state. Numerical values of invariants
and entanglement monotones for states known or conjectured to be maximally
entangled four qubit states are reported and nature of quantum correlations
in these states analyzed in section V followed by a summary of results in
section VI.

\section{Definition of a $K-$way negativity font}

Consider a bipartite system consisting of two distinguishable subsystems $A$
and $B$, each of arbitrary dimension, in pure state $\widehat{\rho }$. The
global negativity\ \cite{zycz98,vida02} of partial transpose $\widehat{\rho }%
_{G}^{T_{A}}$ (partial transpose with respect to $A$) is defined as%
\begin{equation}
N_{G}^{A}=\frac{1}{d_{A}-1}\left( \left\Vert \rho _{G}^{T_{A}}\right\Vert
_{1}-1\right) ,  \label{negdef}
\end{equation}%
where $\left\Vert \widehat{\rho }\right\Vert _{1}$ is the trace norm of $%
\widehat{\rho }$. A general $N-$qubit pure state reads as%
\begin{equation}
\left\vert \Psi ^{A_{1}A_{2}...A_{N}}\right\rangle
=\sum_{i_{1}i_{2}...i_{N}}a_{i_{1}i_{2}...i_{N}}\left\vert
i_{1}i_{2}...i_{N}\right\rangle \qquad \widehat{\rho }=\left\vert \Psi
^{A_{1}A_{2}...A_{N}}\right\rangle \left\langle \Psi
^{A_{1}A_{2}...A_{N}}\right\vert ,  \label{nqubit}
\end{equation}%
where $\left\vert i_{1}i_{2}...i_{N}\right\rangle $ are the basis vectors
spanning $2^{N}$ dimensional Hilbert space and $A_{p}$ is the location of
qubit $p$ ($p=1$ to $N$). The coefficients $a_{i_{1}i_{2}...i_{N}}$ are
complex numbers. The basis states of a single qubit are labelled by $i_{m}=0$
and $1,$ where $m=1,...,N$. The matrix elements of global partial transpose $%
\widehat{\rho }_{G}^{T_{p}}$ with respect to qubit $p$ are obtained from $%
\widehat{\rho }$ through 
\begin{equation}
\left\langle i_{1}i_{2}...i_{N}\right\vert \widehat{\rho }%
_{G}^{T_{p}}\left\vert j_{1}j_{2}...j_{N}\right\rangle =\left\langle
i_{1}i_{2}...i_{p-1}j_{p}i_{p+1}...i_{N}\right\vert \widehat{\rho }%
\left\vert j_{1}j_{2}...j_{p-1}i_{p}j_{p+1}...j_{N}\right\rangle .
\label{gpt}
\end{equation}%
Peres PPT separability criterion \cite{pere96} states that the partial
transpose $\widehat{\rho }_{G}^{T_{p}}$ of a separable state is positive.

Rewrite $N-$qubit pure state as $\left\vert \Psi
^{A_{1}A_{2}...A_{N}}\right\rangle
=\sum\limits_{i_{3}i_{4}...i_{N}}\left\vert F\right\rangle
_{00i_{3}i_{4}...i_{N}}$, where%
\begin{eqnarray}
\left\vert F\right\rangle _{00i_{3}i_{4}...i_{N}}
&=&a_{00i_{3}i_{4}...i_{N}}\left\vert 00i_{3}i_{4}...i_{N}\right\rangle
+a_{10i_{3}i_{4}...i_{N}}\left\vert 10i_{3}i_{4}...i_{N}\right\rangle  \notag
\\
&&+a_{01i_{3}+1i_{4}+1...i_{N}+1}\left\vert
01i_{3}+1i_{4}+1...i_{N}+1\right\rangle  \notag \\
&&+a_{11i_{3}+1i_{4}+1...i_{N}+1}\left\vert
11i_{3}+1i_{4}+1...i_{N}+1\right\rangle .
\end{eqnarray}%
Here $i_{m}+1=0$ for $i_{m}=1$ and $i_{m}+1=1$ for $i_{m}=0$. The
entanglement of $\chi ^{00i_{3}i_{4}...i_{N}}=$ $\left\vert F\right\rangle
_{00i_{3}i_{4}...i_{N}}\left\langle F\right\vert $ is quantified by%
\begin{equation}
\left( N_{G}^{A_{1}}(\chi ^{00i_{3}i_{4}...i_{N}})\right) ^{2}=4\left\vert
\det \left[ 
\begin{array}{cc}
a_{00i_{3}i_{4}...i_{N}} & a_{01i_{3}+1i_{4}+1...i_{N}+1} \\ 
a_{10i_{3}i_{4}...i_{N}} & a_{11i_{3}+1i_{4}+1...i_{N}+1}%
\end{array}%
\right] \right\vert ^{2}=4\left\vert D^{00i_{3}i_{4}...i_{N}}\right\vert
^{2}.
\end{equation}%
Since determinant $D^{00i_{3}i_{4}...i_{N}}=\det \nu
_{N}^{00i_{3}i_{4}...i_{N}}$ determines $N_{G}^{A_{1}}(\chi
^{00i_{3}i_{4}...i_{N}}),$ we refer to $2\times 2$ matrix of probability
amplitudes 
\begin{equation}
\nu _{N}^{00i_{3}i_{4}...i_{N}}=\left[ 
\begin{array}{cc}
a_{00i_{3}i_{4}...i_{N}} & a_{01i_{3}+1i_{4}+1...i_{N}+1} \\ 
a_{10i_{3}i_{4}...i_{N}} & a_{11i_{3}+1i_{4}+1...i_{N}+1}%
\end{array}%
\right] ,  \label{nwayfont}
\end{equation}%
as a negativity font of $N-$way entanglement in $\left\vert \Psi
^{A_{1},A_{2},...A_{N}}\right\rangle $.

In general, if $\widehat{\rho }$ is a pure state, then the negative
eigenvalue of $4\times 4$ sub-matrix of global partial transpose $\widehat{%
\rho }_{G}^{T_{p}}$or a $K-$way partial transpose $\widehat{\rho }%
_{K}^{T_{p}}$ \cite{shar09} in the space spanned by distinct basis vectors $%
\left\vert i_{1}i_{2}...i_{p}...i_{N}\right\rangle $, $\left\vert
j_{1}j_{2}...j_{p}=i_{p}+1...j_{N}\right\rangle $, $\left\vert
i_{1}i_{2}...j_{p}...i_{N}\right\rangle $, and $\left\vert
j_{1}j_{2}...i_{p}...j_{N}\right\rangle $ is \ $\lambda ^{-}=-\left\vert
\det \left( \nu _{K}^{i_{1}i_{2}...i_{p}...i_{N}}\right) \right\vert $ with $%
\nu _{K}^{i_{1}i_{2}...i_{p}...i_{N}}$ defined as%
\begin{equation}
\nu _{K}^{i_{1}i_{2}...i_{p}...i_{N}}=\left[ 
\begin{array}{cc}
a_{i_{1}i_{2}...i_{p}...i_{N}} & a_{j_{1}j_{2}...i_{p}...j_{N}} \\ 
a_{i_{1}i_{2}...j_{p}=i_{p}+1...i_{N}} & 
a_{j_{1}j_{2}...j_{p}=i_{p}+1...j_{N}}%
\end{array}%
\right] ,  \label{kwayfont}
\end{equation}%
where $K=\sum\limits_{m=1}^{N}(1-\delta _{i_{m},j_{m}})$ $\left( 2\leq K\leq
N\right) $. Here $\delta _{i_{m},j_{m}}=1$ for $i_{m}=j_{m}$, and $\delta
_{i_{m},j_{m}}=0$ for $i_{m}\neq j_{m}$. The $2\times 2$ matrix $\nu
_{K}^{i_{1}i_{2}...i_{p}...i_{N}}$ defines a $K-$way negativity font. To
distinguish between different $K-$way negativity fonts we shall replace
subscript $K$ in Eq. (\ref{kwayfont}) by a list of qubit states for which $%
\delta _{i_{m},j_{m}}=1$. In other words a $K-$way font involving qubits $%
A_{q+1}$ to $A_{q+K}$ such that $\sum\limits_{m=1}^{N}(1-\delta
_{i_{m},j_{m}})=\sum\limits_{m=q+1}^{q+K}(1-\delta _{i_{m},j_{m}})=K$, reads
as%
\begin{eqnarray}
&&\nu _{\left( A_{1}\right) _{i_{1}},\left( A_{2}\right) _{i_{2}},...\left(
A_{q}\right) _{i_{q}}\left( A_{q+K+1}\right) _{i_{q+K+1}}...\left(
A_{N}\right) _{i_{N}}}^{i_{1}i_{2}...i_{p}...i_{N}}  \notag \\
&=&\left[ 
\begin{array}{cc}
a_{i_{1}i_{2}...i_{p}...i_{N}} & 
a_{i_{1}i_{2}...i_{q},i_{q+1}+1,i_{q+2}+1...i_{p}...,i_{q+K-1}+1,i_{q+K}+1,i_{q+K+1},...i_{N}}
\\ 
a_{i_{1}i_{2}...i_{p}+1...i_{N}} & 
a_{i_{1}i_{2}...i_{q},i_{q+1}+1,i_{q+2}+1...i_{p}+1...,i_{q+K-1}+1,i_{q+K}+1,i_{q+K+1},...i_{N}}%
\end{array}%
\right] ,
\end{eqnarray}%
and its determinant is represented by 
\begin{eqnarray}
&&D_{\left( A_{1}\right) _{i_{1}},\left( A_{2}\right) _{i_{2}},...\left(
A_{q}\right) _{i_{q}}\left( A_{q+K+1}\right) _{i_{q+K+1}}...\left(
A_{N}\right) _{i_{N}}}^{i_{q+1}...i_{p}...i_{q+k-1}i_{q+k}}  \notag \\
&=&\det \left( \nu _{\left( A_{1}\right) _{i_{1}},\left( A_{2}\right)
_{i_{2}},...\left( A_{q}\right) _{i_{q}}\left( A_{q+K+1}\right)
_{i_{q+K+1}}...\left( A_{N}\right)
_{i_{N}}}^{i_{1}i_{2}...i_{p}...i_{N}}\right) .
\end{eqnarray}%
Thus the determinant of a $K-$way font in an $N$ qubit state has $N-K$
subscripts and $K$ superscripts. In this notation no subscript is needed for
determinant of an $N-$way negativity font. The general rule to represent the
determinants of negativity fonts is that the qubit states are ordered
according to the location of the qubits with the states that appear in the
subscript not being present in the superscript. One can identify the
determinants of negativity fonts with Pl\"{u}cker coordinates in ref. \cite%
{heyd06}, where Pl\"{u}cker coordinate equations of Grassmann variety have
been used to construct entanglement monotones for multi-qubit states.

\subsection{Negativity fonts in K-way partial transpose}

To construct a $K-$way partially transposed matrix \cite{shar09} from the
state operator $\widehat{\rho }$, every matrix element $\left\langle
i_{1}i_{2}...i_{N}\right\vert \widehat{\rho }\left\vert
j_{1}j_{2}...j_{N}\right\rangle $ is labelled by a number $%
K=\sum\limits_{m=1}^{N}(1-\delta _{i_{m},j_{m}})$. The $K-$way partial
transpose ($K>2$) of\ $\rho $ with respect to subsystem $p$ is obtained by
selective transposition such that%
\begin{eqnarray}
\left\langle i_{1}i_{2}...i_{N}\right\vert \widehat{\rho }%
_{K}^{T_{p}}\left\vert j_{1}j_{2}...j_{N}\right\rangle &=&\left\langle
i_{1}i_{2}...i_{p-1}j_{p}i_{p+1}...i_{N}\right\vert \widehat{\rho }%
\left\vert j_{1}j_{2}...j_{p-1}i_{p}j_{p+1}...j_{N}\right\rangle ,  \notag \\
\text{if}\quad \sum\limits_{m=1}^{N}(1-\delta _{i_{m},j_{m}}) &=&K,\quad 
\text{and }\quad \delta _{i_{p},j_{p}}=0  \label{ptk1}
\end{eqnarray}%
and%
\begin{eqnarray}
\left\langle i_{1}i_{2}...i_{N}\right\vert \widehat{\rho }%
_{K}^{T_{p}}\left\vert j_{1}j_{2}...j_{N}\right\rangle &=&\left\langle
i_{1}i_{2}...i_{N}\right\vert \widehat{\rho }\left\vert
j_{1}j_{2}...j_{N}\right\rangle ,  \notag \\
\quad \text{if}\quad \sum\limits_{m=1}^{N}(1-\delta _{i_{m},j_{m}}) &\neq &K.
\label{ptk2}
\end{eqnarray}%
while%
\begin{eqnarray}
\left\langle i_{1}i_{2}...i_{N}\right\vert \widehat{\rho }%
_{2}^{T_{p}}\left\vert j_{1}j_{2}...j_{N}\right\rangle &=&\left\langle
i_{1}i_{2}...i_{p-1}j_{p}i_{p+1}...i_{N}\right\vert \widehat{\rho }%
\left\vert j_{1}j_{2}...j_{p-1}i_{p}j_{p+1}...j_{N}\right\rangle ,  \notag \\
\quad \text{if}\quad \sum\limits_{m=1}^{N}(1-\delta _{i_{m},j_{m}}) &=&1%
\text{ or }2,\quad \text{and }\quad \delta _{i_{p},j_{p}}=0  \label{pt21}
\end{eqnarray}%
and%
\begin{eqnarray}
\left\langle i_{1}i_{2}...i_{N}\right\vert \widehat{\rho }%
_{2}^{T_{p}}\left\vert j_{1}j_{2}...j_{N}\right\rangle &=&\left\langle
i_{1}i_{2}...i_{N}\right\vert \widehat{\rho }\left\vert
j_{1}j_{2}...j_{N}\right\rangle ,  \notag \\
\quad \text{if}\quad \sum\limits_{m=1}^{N}(1-\delta _{i_{m},j_{m}}) &\neq &1%
\text{ or }2.  \label{pt22}
\end{eqnarray}%
The $K-$way negativity calculated from $K-$way partial transpose of matrix $%
\rho $ with respect to subsystem $p$, is defined as $N_{K}^{A_{p}}=\left(
\left\Vert \rho _{K}^{T_{p}}\right\Vert _{1}-1\right) $. Using the
definition of trace norm and the fact that $tr(\rho _{K}^{T_{p}})=1$, we get 
$N_{K}^{A_{p}}=2\sum_{i}\left\vert \lambda _{i}^{K-}\right\vert $, $\lambda
_{i}^{K-}$ being the negative eigenvalues of matrix $\rho _{K}^{T_{p}}$. The 
$K-$way negativity ($2\leq K\leq N)$, defined as the negativity of $K-$way
partial transpose, is determined by the presence or absence of $K-$way
quantum coherences in the composite system. By $K-$way coherences we mean
the type of coherences present in a $K-$qubit GHZ- like state. The
negativity $N_{K}^{A_{p}}$ is a measure of all possible types of
entanglement attributed to $K-$ way coherences. It was shown in refs. \cite%
{shar07,shar08,shar09} that the global partial transpose of an $N-$qubit
state may be written as a sum of \ $K-$way partial transposes $\left( 2\leq
K\leq N\right) $ that is 
\begin{equation}
\widehat{\rho }_{G}^{T_{p}}=\sum\limits_{K=2}^{N}\widehat{\rho }%
_{K}^{T_{p}}-(N-2)\widehat{\rho }.  \label{3n}
\end{equation}%
By rewriting the global partial transpose as a sum of $K-$way partial
transposes, the negativity fonts are distributed amongst $N-1$ partial
transposes. Contributions of partial transposes to global negativity,
referred to as partial $K-$way negativities are not unitary invariants, but
their values coincide with those of three tangle and concurrences for three
qubit canonical state\cite{shar07}.

\section{Transformation equations for determinants of negativity fonts,
global negativity and two-qubit invariants}

To derive expressions for LU invariants which measure genuine $N-$body
quantum correlations present in the state, the transformation equations
under LU are written, for negativity fonts characterizing the $N-$way
partial transpose and $\left( N-1\right) $ way partial transpose. Two qubit
invariants obtained from transformation equations pave the way to
construction of $N-$qubit LU invariants to be used to write the entanglement
monotones. In the following, an invariant named $\mathcal{I}$ represented by 
$\left( \mathcal{I}_{K}\right) _{\left( A_{x+1}\right) _{i_{x+1}}...\left(
A_{N}\right) _{i_{N}}}^{A_{1}...A_{x}}$, is understood to be invariant under
the action of local unitaries on qubits $A_{1}$, $A_{2}$,$...,A_{x}$ of the
N qubit system. In general, the superscript outside the bracket will list
the qubits in the subsystem of which $\mathcal{I}_{K}$ is an invariant,
while subscript lists the remaining qubits and their states. In case no
state specification is needed, subscript is redunant as such will not be
written. When $\left( \mathcal{I}_{K}\right) $ is an N-qubit invariant both
sub and superscripts are redundant and will not be posted. Subscript $K$ in $%
\mathcal{I}_{K}$ indicates that by suitable choice of local unitaries the
invariant can be expressed in terms of determinants of $K-$way negativity
fonts. Determinant of an $N-$way negativity font%
\begin{equation}
D^{i_{1}i_{2}...i_{p}=0...i_{N}}=\det \left[ 
\begin{array}{cc}
a_{i_{1}i_{2}...i_{p}=0...i_{N}} & a_{i_{1}+1,i_{2}+1,...i_{p}=0...i_{N}+1}
\\ 
a_{i_{1}i_{2}...i_{p}=1...i_{N}} & a_{i_{1}+1,i_{2}+1,...i_{p}=1...i_{N}+1}%
\end{array}%
\right] ,
\end{equation}%
is an invariant of U$^{A_{p}}$. Local unitary $U^{A_{q}}=\frac{1}{\sqrt{%
1+\left\vert x\right\vert ^{2}}}\left[ 
\begin{array}{cc}
1 & -x^{\ast } \\ 
x & 1%
\end{array}%
\right] $ on qubit $A_{q}$ with $q\neq p$, on the other hand, yields four
transformation equations%
\begin{eqnarray}
\left( D^{i_{1}i_{2}...i_{p}=0,i_{q}=0,...i_{N}}\right) ^{\prime \prime } &=&%
\frac{1}{1+\left\vert x\right\vert ^{2}}\left[
D^{i_{1}i_{2}...i_{p}=0,i_{q}=0...i_{N}}-\left\vert x\right\vert
^{2}D^{i_{1}i_{2}...i_{p}=0,i_{q}=1...i_{N}}\right.   \notag \\
&&\left. +xD_{\left( A_{q}\right)
_{0}}^{i_{1}i_{2}...i_{p}=0...,i_{q-1},i_{q+1},...i_{N}}-x^{\ast }D_{\left(
A_{q}\right) _{1}}^{i_{1}i_{2}...i_{p}=0...i_{q-1}i_{q+1}...i_{N}}\right] 
\label{t1}
\end{eqnarray}%
\begin{eqnarray}
\left( D^{i_{1}i_{2}...i_{p}=0,i_{q}=1,...i_{N}}\right) ^{\prime \prime } &=&%
\frac{1}{1+\left\vert x\right\vert ^{2}}\left[
D^{i_{1}i_{2}...i_{p}=0,i_{q}=1,...i_{N}}-\left\vert x\right\vert
^{2}D^{i_{1}i_{2}...i_{p}=0,i_{q}=0...i_{N}}\right.   \notag \\
&&\left. +xD_{\left( A_{q}\right)
_{0}}^{i_{1}i_{2}...i_{p}=0...,i_{q-1},i_{q+1},...i_{N}}-x^{\ast }D_{\left(
A_{q}\right) _{1}}^{i_{1}i_{2}...i_{p}=0...i_{q-1}i_{q+1}...i_{N}}\right] 
\label{t2}
\end{eqnarray}%
\begin{eqnarray}
\left( D_{\left( A_{q}\right)
_{0}}^{i_{1}i_{2}...i_{p}=0...i_{q-1},i_{q+1}...i_{N}}\right) ^{\prime
\prime } &=&\frac{1}{1+\left\vert x\right\vert ^{2}}\left[ D_{\left(
A_{q}\right) _{0}}^{i_{1}i_{2}...i_{p}=0...,i_{q-1},i_{q+1}...i_{N}}+\left(
x^{\ast }\right) ^{2}D_{\left( A_{q}\right)
_{1}}^{i_{1}i_{2}...i_{p}=0...,i_{q-1},i_{q+1}...i_{N}}\right.   \notag \\
&&\left. -x^{\ast }\left(
D^{i_{1}i_{2}...i_{p}=0,i_{q}=0...i_{N}}+D^{i_{1}i_{2}...i_{p}=0,i_{q}=1...i_{N}}\right) 
\right]   \label{t3}
\end{eqnarray}%
\begin{eqnarray}
\left( D_{\left( A_{q}\right)
_{1}}^{i_{1}i_{2}...i_{p}=0...,i_{q-1},i_{q+1},...i_{N}}\right) ^{\prime
\prime } &=&\frac{1}{1+\left\vert x\right\vert ^{2}}\left[ D_{\left(
A_{q}\right)
_{1}}^{i_{1}i_{2}...i_{p}=0...,i_{q-1},i_{q+1},...i_{N}}+x^{2}D_{\left(
A_{q}\right) _{0}}^{i_{1}i_{2}...i_{p}=0...,i_{q-1},i_{q+1}...i_{N}}\right. 
\notag \\
&&\left. +x\left(
D^{i_{1}i_{2}...i_{p}=0,i_{q}=0...i_{N}}+D^{i_{1}i_{2}...i_{p}=0,i_{q}=1...i_{N}}\right) 
\right]   \label{t4}
\end{eqnarray}%
relating $N-$way and $\left( N-1\right) -$way negativity fonts. Eliminating
variable $x$, invariants of $U^{A_{p}}U^{A_{q}}$ are found to be 
\begin{eqnarray}
\left( M_{N}\right) ^{A_{p}A_{q}} &=&\left\vert \left(
D^{i_{1}i_{2}...i_{p}=0,i_{q}=0,...i_{N}}\right) ^{\prime \prime
}\right\vert ^{2}+\left\vert \left(
D^{i_{1}i_{2}...i_{p}=0,i_{q}=1,...i_{N}}\right) ^{\prime \prime
}\right\vert ^{2}  \notag \\
&&+\left\vert \left( D_{\left( A_{q}\right)
_{0}}^{i_{1}i_{2}...i_{p}=0...i_{q-1}i_{q+1}...i_{N}}\right) ^{\prime \prime
}\right\vert ^{2}+\left\vert \left( D_{\left( A_{q}\right)
_{1}}^{i_{1}i_{2}...i_{p}=0...i_{q-1}i_{q+1}...i_{N}}\right) ^{\prime \prime
}\right\vert ^{2}  \notag \\
&=&\left\vert \left( D^{i_{1}i_{2}...i_{p}=0,i_{q}=0,...i_{N}}\right)
\right\vert ^{2}+\left\vert \left(
D^{i_{1}i_{2}...i_{p}=0,i_{q}=1,...i_{N}}\right) \right\vert ^{2}  \notag \\
&&+\left\vert D_{\left( A_{q}\right)
_{0}}^{i_{1}i_{2}...i_{p}=0...,i_{q-1},i_{q+1},...i_{N}}\right\vert
^{2}+\left\vert D_{\left( A_{q}\right)
_{1}}^{i_{1}i_{2}...i_{p}=0...i_{q-1}i_{q+1}...i_{N}}\right\vert ^{2},
\label{d0mod}
\end{eqnarray}%
which is real, a degree two invariant%
\begin{eqnarray}
\left( T_{N}\right) ^{A_{p}A_{q}} &=&\left(
D^{i_{1}i_{2}...i_{p}=0,i_{q}=0,...i_{N}}\right) ^{\prime \prime }-\left(
D^{i_{1}i_{2}...i_{p}=0,i_{q}=1,...i_{N}}\right) ^{\prime \prime }  \notag \\
&=&D^{i_{1}i_{2}...i_{p}=0i_{q}=0...i_{N}}-D^{i_{1}i_{2}...i_{p}=0i_{q}=1...i_{N}},
\label{d1dif}
\end{eqnarray}%
a degree four invariant%
\begin{eqnarray}
\left( I_{N}\right) ^{A_{p}A_{q}} &=&\left(
D^{i_{1}i_{2}...i_{p}=0i_{q}=0...i_{N}}+D^{i_{1}i_{2}...i_{p}=0i_{q}=1...i_{N}}\right) ^{2}
\notag \\
&&-4D_{\left( A_{q}\right)
_{0}}^{i_{1}i_{2}...i_{p}=0...i_{q-1},i_{q+1}...i_{N}}D_{\left( A_{q}\right)
_{1}}^{i_{1}i_{2}...i_{p}=0...,i_{q-1},i_{q+1},...i_{N}}  \notag \\
&=&\left( \left( D^{i_{1}i_{2}...i_{p}=0,i_{q}=0,...i_{N}}\right) ^{\prime
\prime }+\left( D^{i_{1}i_{2}...i_{p}=0,i_{q}=1,...i_{N}}\right) ^{\prime
\prime }\right) ^{2}  \notag \\
&&-4\left( D_{\left( A_{q}\right)
_{0}}^{i_{1}i_{2}...i_{p}=0...i_{q-1},i_{q+1}...i_{N}}\right) ^{\prime
\prime }\left( D_{\left( A_{q}\right)
_{1}}^{i_{1}i_{2}...i_{p}=0...,i_{q-1},i_{q+1},...i_{N}}\right) ^{\prime
\prime },  \label{d2sum}
\end{eqnarray}%
and combining Eqs. (\ref{d1dif}) and (\ref{d2sum}), we obtain%
\begin{eqnarray}
\left( P_{N}\right) ^{A_{p}A_{q}}
&=&D^{i_{1}i_{2}...i_{p}=0i_{q}=0...i_{N}}D^{i_{1}i_{2}...i_{p}=0i_{q}=1...i_{N}}
\notag \\
&&-D_{\left( A_{q}\right)
_{0}}^{i_{1}i_{2}...i_{p}=0...i_{q-1},i_{q+1}...i_{N}}D_{\left( A_{q}\right)
_{1}}^{i_{1}i_{2}...i_{p}=0...,i_{q-1},i_{q+1},...i_{N}}  \notag \\
&=&\left( D^{i_{1}i_{2}...i_{p}=0,i_{q}=0,...i_{N}}\right) ^{\prime \prime
}\left( D^{i_{1}i_{2}...i_{p}=0,i_{q}=1,...i_{N}}\right) ^{\prime \prime } 
\notag \\
&&-\left( D_{\left( A_{q}\right)
_{0}}^{i_{1}i_{2}...i_{p}=0...i_{q-1},i_{q+1}...i_{N}}\right) ^{\prime
\prime }\left( D_{\left( A_{q}\right)
_{1}}^{i_{1}i_{2}...i_{p}=0...,i_{q-1},i_{q+1},...i_{N}}\right) ^{\prime
\prime }.  \label{d3prod}
\end{eqnarray}%
Similarly the differences $\left( M_{N}\right) ^{A_{p}A_{q}}-\left\vert
\left( I_{N}\right) ^{A_{p}A_{q}}\right\vert $ and $\left( M_{N}\right)
^{A_{p}A_{q}}-\left\vert \left( T_{N}\right) ^{A_{p}A_{q}}\right\vert ^{2}$
are useful to write down different $N-$qubit invariants in alternate forms.

Transformation equations under LU for determinants of negativity fonts
characterizing $K-$way partial transpose and $\left( K-1\right) $ way
partial transpose with $K<N$, yield two qubit invariants $\left(
M_{K}\right) ^{A_{p}A_{q}}$, $\left( I_{K}\right) ^{A_{p}A_{q}}$, $\left(
T_{K}\right) ^{A_{p}A_{q}}$, and $\left( P_{K}\right) ^{A_{p}A_{q}}$
analogous to $N-$way case.

\subsection{Global negativity and negativity fonts}

It follows from Eq. (\ref{d0mod}) that \ by summing up the squared moduli of
determinants of all negativity fonts in a partial transpose we obtain an
N-qubit invariant. Recalling that the maximum value that modulus of
determinant of a single negativity font may have is $\frac{1}{2}$,
multiplying the invariant by four leads to an invariant with maximum value
equal to one. Next, the relation between global negativity and linear
entropy of reduced single qubit state is used to demonstrate that the
invariant obtained is nothing but the global negativity defined as in Eq. (%
\ref{negdef}).

Linear entropy, defined as%
\begin{equation}
S=\frac{d_{A}}{d_{A}-1}\left( 1-Tr\left( \rho ^{A}\right) ^{2}\right)
\end{equation}%
measures the purity of state $\rho ^{A}=Tr_{B}$ $\left( \widehat{\rho }%
\right) $ and also detects bipartite entanglement of subsystems $A$ with $B$%
. If $A=A_{p}$, the $\left( p^{th}\right) $ qubit of an $N-$qubit quantum
system, then squared negativity $\left( N_{G}^{A_{p}}\right) ^{2}$ is known
to be equal to linear entropy of single qubit reduced state $\widehat{\rho }%
^{A_{p}}=tr_{A_{1}...A_{p-1}A_{p+1}...A_{N}}\left( \widehat{\rho }\right) $
that is%
\begin{equation}
\left( N_{G}^{A_{p}}\right) ^{2}=2\left( 1-tr\left[ \left( \widehat{\rho }%
^{A_{p}}\right) ^{2}\right] \right) .  \label{negred}
\end{equation}%
Choosing $p=1$, we write the pure state as $\widehat{\rho }%
=\sum\limits_{I,J}\rho _{i_{1}Ij_{1}J}\left\vert i_{1}I\right\rangle
\left\langle j_{1}J\right\vert $, where $I=\sum\limits_{m=2}^{N}i_{m}2^{m-1}$
labels the $\left( N-1\right) $ qubit state sans qubit $A_{1}$. Using Eq. (%
\ref{negred}) and $tr\left( \widehat{\rho }^{A_{1}}\right) =1,$ we obtain%
\begin{equation}
\left( N_{G}^{A_{1}}\right) ^{2}=4\sum\limits_{I,J}\left( \rho _{1I0I}\rho
_{0J1J}-\rho _{0I0I}\rho _{1J1J}\right) .
\end{equation}%
Next defining $L=\sum\limits_{\substack{ m=3  \\ m\neq p}}^{N}i_{m}2^{m-1}$
and $M=\sum\limits_{\substack{ m=3  \\ m\neq p}}^{N}j_{m}2^{m-1}$, expansion
of $\left( N_{G}^{A_{1}}\right) ^{2}$ reads as%
\begin{eqnarray}
\left( N_{G}^{A_{1}}\right) ^{2} &=&4\sum\limits_{L,M}\left( \rho
_{10L00L}\rho _{00M10M}-\rho _{00L00L}\rho _{10M10M}\right)  \notag \\
&&+4\sum\limits_{L,M}\left( \rho _{10L00L}\rho _{01M11M}-\rho _{00L00L}\rho
_{11M11M}\right)  \notag \\
&&+4\sum\limits_{L,M}\left( \rho _{11L,01L}\rho _{00M10M}-\rho
_{01L,01L}\rho _{10M10M}\right)  \notag \\
&&+4\sum\limits_{L,M}\left( \rho _{11L,01L}\rho _{01M11M}-\rho
_{01L,01L}\rho _{11M11M}\right)
\end{eqnarray}%
which in terms of probability amplitudes has the form%
\begin{equation}
\left( N_{G}^{A_{1}}\right) ^{2}=4\sum\limits_{L,M}\left\vert \left(
a_{00L}a_{11M}-a_{10L}a_{01M}\right) \right\vert ^{2}
\end{equation}%
After identifying the determinant $\left(
a_{00L}a_{11M}-a_{10L}a_{01M}\right) $ with 
\begin{equation}
\det \nu _{K}^{00L}\equiv \det \left[ 
\begin{array}{cc}
a_{00i_{3}...i_{N}} & a_{01j_{3}...j_{N}} \\ 
a_{10i_{3}...i_{N}} & a_{11j_{3}...j_{N}}%
\end{array}%
\right] ,
\end{equation}%
that is the determinant of a $K-$way negativity font, the squared
negativity\ is expressed in terms of determinants of all negativity fonts in 
$\widehat{\rho }_{G}^{T_{1}}$as 
\begin{equation}
\left( N_{G}^{A_{1}}\right) ^{2}=4\sum\limits_{L,K=2\text{ to }N}\left\vert
\det \nu _{K}^{00L}\right\vert ^{2}.  \label{negativity}
\end{equation}%
Global negativity arising due to all the negativity fonts present in $%
\widehat{\rho }_{G}^{T_{p}}$ measures the entanglement of qubit $p$ with
it's complement and is known to be an entanglement monotone \cite{vida02}.

\section{Four qubit invariants}

For $N=4$, with determinants of four-way negativity fonts defined as 
\begin{equation}
D^{00i_{3}i_{4}}=\det \left( 
\begin{array}{cc}
a_{00i_{3}i_{4}} & a_{01i_{3}+1,i_{4}+1} \\ 
a_{10i_{3}i_{4}} & a_{11i_{3}+1,i_{4}+1}%
\end{array}%
\right) ,
\end{equation}%
four qubit pure state invariant with negativity fonts lying solely in
four-way partial transpose is given by%
\begin{equation}
T_{4}=D^{0000}+D^{0011}-D^{0010}-D^{0001}.  \label{4-invariant}
\end{equation}%
Invariant $T_{4}$ is identified with degree two invariant H of ref. \cite%
{luqu03} which is also one of the hyperdeterminants of Cayley. A four qubit
state having quantum correlations of the type present in a four qubit GHZ
state, is distinguished from other states by a non zero $T_{4}$. These
quantum correlations are lost without leaving any residue, on the loss of a
single qubit and are a collective property of four qubit state. It is known 
\cite{luqu03} that four tangle defined as 
\begin{equation}
\tau _{4}=4\left\vert \left( D^{0000}+D^{0011}-D^{0010}-D^{0001}\right)
^{2}\right\vert ,  \label{tau4}
\end{equation}%
by itself is not enough to detect four qubit genuine entanglement, being
non-zero for the product of entangled two qubit states in which case
invariants of higher degree are needed to detect GHZ\ like entanglement.

Local unitary transformations may be used to concentrate the negativity
fonts on a selected $\rho _{K}^{T_{p}}$ in the expansion of $\rho
_{G}^{T_{p}} $ given by Eq. (\ref{3n}). When $\rho _{G}^{T_{p}}=$ $\rho
_{4}^{T_{p}}$ and $\tau _{4}\neq 0$, we have a GHZ like four qubit state.
Four qubit states with each qubit entangled to at least one qubit and $\tau
_{4}\neq 0$, can have canonical states with 
\begin{eqnarray*}
\rho _{G}^{T_{p}} &=&\rho _{4}^{T_{p}}+\rho _{3}^{T_{p}}+\rho
_{2}^{T_{p}}-2\rho ,\quad \rho _{G}^{T_{p}}=\rho _{4}^{T_{p}}+\rho
_{3}^{T_{p}}-\rho , \\
\rho _{G}^{T_{p}} &=&\rho _{4}^{T_{p}}+\rho _{2}^{T_{p}}-\rho ,\quad \rho
_{G}^{T_{p}}=\rho _{4}^{T_{p}}.
\end{eqnarray*}%
The class with $\tau _{4}=0$, allows for two equivalent canonical state
descriptions that is 
\begin{equation*}
\rho _{G}^{T_{p}}=\rho _{4}^{T_{p}}+\rho _{2}^{T_{p}}-\rho ,\quad \text{or}%
\quad \rho _{G}^{T_{p}}=\rho _{3}^{T_{p}}+\rho _{2}^{T_{p}}-\rho .
\end{equation*}%
Therefore the difference 
\begin{equation}
\Delta _{4}=\sum\limits_{p=1}^{4}\left( N_{G}^{A_{p}}\right) ^{2}-\tau _{4},
\end{equation}%
for four qubit pure state may be taken as a measure of three-way plus
two-way coherences.

\subsection{Entanglement of two and three qubits in Four qubit states}

As mentioned before, to distinguish between the product of two qubit
entangled states with $\tau _{4}\neq 0$ and states with all four qubits
entangled to each other we need additional invariants. In ref. \cite{shar101}%
, along with the degree two invariant of Eq. (\ref{4-invariant}), we
reported three degree four invariants that detect quantum correlations in a
four qubit state. In this section, we list those invariants and identify two
distinct types of three qubit invariants that constitute a four qubit
invariant. Three-way and two-way negativity font determinants for four
qubits are defined as%
\begin{eqnarray}
\qquad D_{\left( A_{2}\right) _{i_{2}}}^{0i_{3}i_{4}} &=&\det \left( 
\begin{array}{cc}
a_{0i_{2}i_{3}i_{4}} & a_{0i_{2}i_{3}+1,i_{4}+1} \\ 
a_{1i_{2}i_{3}i_{4}} & a_{1i_{2}i_{3}+1,i_{4}+1}%
\end{array}%
\right) ,\qquad D_{\left( A_{3}\right) _{i_{3}}}^{0i_{2}i_{4}}=\det \left( 
\begin{array}{cc}
a_{00i_{3}i_{4}} & a_{01i_{3},i_{4}+1} \\ 
a_{10i_{3}i_{4}} & a_{11i_{3},i_{4}+1}%
\end{array}%
\right) ,  \notag \\
\qquad D_{\left( A_{4}\right) _{i_{4}}}^{0i_{2}i_{3}} &=&\det \left( 
\begin{array}{cc}
a_{00i_{3}i_{4}} & a_{01i_{3}+1,i_{4}} \\ 
a_{10i_{3}i_{4}} & a_{10i_{3}+1,i_{4}}%
\end{array}%
\right) ,\qquad D_{\left( A_{p}\right) _{i_{p}}\left( A_{q}\right)
_{i_{q}}}^{00}=\det \left( \nu _{\left( A_{p}\right) _{i_{p}}\left(
A_{q}\right) _{i_{q}}}^{00i_{p}i_{q}}\right) .
\end{eqnarray}%
Using Eq. (\ref{d1dif})) for four qubits and identifying the terms 
\begin{equation*}
D^{0000}-D^{0001}+D^{0010}-D^{0011},
\end{equation*}%
\begin{equation*}
\left( D_{\left( A_{3}\right) _{0}}^{000}-D_{\left( A_{3}\right)
_{0}}^{001}\right) \times \left( D_{\left( A_{3}\right)
_{1}}^{000}-D_{\left( A_{3}\right) _{1}}^{001}\right) ,\left( D_{\left(
A_{2}\right) _{0}}^{000}-D_{\left( A_{2}\right) _{0}}^{001}\right) \times
\left( D_{\left( A_{2}\right) _{1}}^{000}-D_{\left( A_{2}\right)
_{1}}^{001}\right) ,
\end{equation*}%
\begin{equation}
\left( D_{\left( A_{2}\right) _{0}\left( A_{3}\right) _{0}}^{00}\right)
\times \left( D_{\left( A_{2}\right) _{1}\left( A_{3}\right)
_{1}}^{00}\right) ,\left( D_{\left( A_{2}\right) _{0}\left( A_{3}\right)
_{1}}^{00}\right) \times \left( D_{\left( A_{2}\right) _{1}\left(
A_{3}\right) _{0}}^{00}\right) ,
\end{equation}%
as invariants of U$^{A_{1}}$U$^{A_{4}},$ application of Eq. (\ref{d2sum}))
leads to four qubit invariant%
\begin{eqnarray}
\left( J_{4}^{A_{1}A_{4}}\right) ^{A_{1}A_{2}A_{3}A_{4}} &=&\left(
D^{0000}-D^{0001}+D^{0010}-D^{0011}\right) ^{2}  \notag \\
&&+8\left( D_{\left( A_{2}\right) _{0}\left( A_{3}\right)
_{0}}^{00}D_{\left( A_{2}\right) _{1}\left( A_{3}\right)
_{1}}^{00}+D_{\left( A_{2}\right) _{0}\left( A_{3}\right)
_{1}}^{00}D_{\left( A_{2}\right) _{1}\left( A_{3}\right) _{0}}^{00}\right)  
\notag \\
&&-4\left( D_{\left( A_{3}\right) _{0}}^{000}-D_{\left( A_{3}\right)
_{0}}^{001}\right) \left( D_{\left( A_{3}\right) _{1}}^{000}-D_{\left(
A_{3}\right) _{1}}^{001}\right)   \notag \\
&&-4\left( D_{\left( A_{2}\right) _{0}}^{000}-D_{\left( A_{2}\right)
_{0}}^{001}\right) \left( D_{\left( A_{2}\right) _{1}}^{000}-D_{\left(
A_{2}\right) _{1}}^{001}\right) .  \label{j14}
\end{eqnarray}%
From the structure of $\left( J_{4}^{A_{1}A_{4}}\right)
^{A_{1}A_{2}A_{3}A_{4}}$ we deduce that four qubit 
\begin{equation*}
\left\vert GHZ\right\rangle =\frac{1}{\sqrt{2}}\left( \left\vert
0000\right\rangle +\left\vert 1111\right\rangle \right) 
\end{equation*}%
state with $\left( J_{4}^{A_{1}A_{4}}\right) ^{A_{1}A_{2}A_{3}A_{4}}=\left(
D^{0000}\right) ^{2}=\frac{1}{4}$ is unitary equivalent to the state 
\begin{eqnarray}
\left\vert 1\right\rangle  &=&\frac{1}{\sqrt{8}}\left( \left\vert
0000\right\rangle +\left\vert 1111\right\rangle +\left\vert
0100\right\rangle -\left\vert 1011\right\rangle \right.   \notag \\
&&\left. +\left\vert 0010\right\rangle -\left\vert 1101\right\rangle
+\left\vert 0110\right\rangle +\left\vert 1001\right\rangle \right) ,  \notag
\end{eqnarray}%
with four-way coherences transformed to three and two way coherences such
that 
\begin{equation*}
\left( D^{0000}-D^{0001}+D^{0010}-D^{0011}\right) ^{2}=0,
\end{equation*}%
and 
\begin{equation*}
\left( J_{4}^{A_{1}A_{4}}\right) ^{A_{1}A_{2}A_{3}A_{4}}=-4\left( D_{\left(
A_{3}\right) _{0}}^{000}-D_{\left( A_{3}\right) _{0}}^{001}\right) \left(
D_{\left( A_{3}\right) _{1}}^{000}-D_{\left( A_{3}\right) _{1}}^{001}\right)
=\frac{1}{4}.
\end{equation*}%
In the present context, $J_{4}^{\left( A_{p}A_{q}\right) }$ are always four
qubit invariants, therefore, the superscript $A_{1}A_{2}A_{3}A_{4}$ will be
understood, from this point on.

To understand the role of three qubit correlations, we rewrite a four qubit
state as 
\begin{equation}
\left\vert \Psi \right\rangle =\left\vert \Phi _{0}\right\rangle \left\vert
0\right\rangle _{A_{3}}+\left\vert \Phi _{1}\right\rangle \left\vert
1\right\rangle _{A_{3}},
\end{equation}%
where 
\begin{equation}
\left\vert \Phi _{0}\right\rangle
=\sum\limits_{i_{1}i_{2}i_{4}}a_{i_{1}i_{2}0i_{4}}\left\vert
i_{1}i_{2}i_{4}\right\rangle ,\quad \left\vert \Phi _{1}\right\rangle
=\sum_{i_{1}i_{2}i_{4}}a_{i_{1}i_{2}1i_{4}}\left\vert
i_{1}i_{2}i_{4}\right\rangle ,
\end{equation}%
are three qubit states characterized by three qubit invariants $\left(
I_{3}\right) _{\left( A_{3}\right) _{0}}^{A_{1}A_{2}A_{4}}$ and $\left(
I_{3}\right) _{\left( A_{3}\right) _{1}}^{A_{1}A_{2}A_{4}}$ with three
tangles given, respectively, by 
\begin{equation}
\left( \tau _{3}\right) _{\left( A_{3}\right) _{0}}=4\left\vert \left(
I_{3}\right) _{\left( A_{3}\right) _{0}}^{A_{1}A_{2}A_{4}}\right\vert
=4\left\vert \left( D_{\left( A_{3}\right) _{0}}^{000}-D_{\left(
A_{3}\right) _{0}}^{001}\right) ^{2}-4D_{\left( A_{2}\right) _{0}\left(
A_{3}\right) _{0}}^{00}D_{\left( A_{2}\right) _{1}\left( A_{3}\right)
_{0}}^{00}\right\vert ,
\end{equation}%
and 
\begin{equation}
\left( \tau _{3}\right) _{\left( A_{3}\right) _{1}}=4\left\vert \left(
I_{3}\right) _{\left( A_{3}\right) _{1}}^{A_{1}A_{2}A_{4}}\right\vert
=4\left\vert \left( D_{\left( A_{3}\right) _{1}}^{000}-D_{\left(
A_{3}\right) _{1}}^{001}\right) ^{2}-4D_{\left( A_{2}\right) _{0}\left(
A_{3}\right) 1}^{00}D_{\left( A_{2}\right) _{1}\left( A_{3}\right)
_{1}}^{00}\right\vert .
\end{equation}%
A polynomial classification scheme in which families of four qubit are
identified through tangle patterns has been suggested recently in \cite%
{vieh11}. We notice that in the context of four qubits, using Eqs. (\ref{t1}-%
\ref{t4}) overall three qubit invariant for qubits $A_{1}A_{2}A_{4}$ may be
written as 
\begin{eqnarray}
\left( I_{3}\right) _{A_{3}}^{A_{1}A_{2}A_{4}} &=&\left( D_{\left(
A_{3}\right) _{0}}^{000}-D_{\left( A_{3}\right) _{0}}^{001}+\left( D_{\left(
A_{3}\right) _{1}}^{000}-D_{\left( A_{3}\right) _{1}}^{001}\right) \right)
^{2}  \notag \\
&&-4\left( D_{\left( A_{2}\right) _{0}\left( A_{3}\right)
_{0}}^{00}+D_{\left( A_{2}\right) _{0}\left( A_{3}\right) _{1}}^{00}\right)
\left( D_{\left( A_{2}\right) _{1}\left( A_{3}\right) _{0}}^{00}+D_{\left(
A_{2}\right) _{1}\left( A_{3}\right) _{1}}^{00}\right) .  \notag \\
&=&\left( I_{3}\right) _{\left( A_{3}\right) _{0}}^{A_{1}A_{2}A_{4}}+\left(
I_{3}\right) _{\left( A_{3}\right) _{1}}^{A_{1}A_{2}A_{4}}+2\left( D_{\left(
A_{3}\right) _{0}}^{000}-D_{\left( A_{3}\right) _{0}}^{001}\right) \left(
D_{\left( A_{3}\right) _{1}}^{000}-D_{\left( A_{3}\right) _{1}}^{001}\right)
\notag \\
&&-4\left( D_{\left( A_{2}\right) _{0}\left( A_{3}\right)
_{0}}^{00}D_{\left( A_{2}\right) _{1}\left( A_{3}\right)
_{1}}^{00}+D_{\left( A_{2}\right) _{0}\left( A_{3}\right)
_{1}}^{00}D_{\left( A_{2}\right) _{1}\left( A_{3}\right) _{0}}^{00}\right) .
\end{eqnarray}%
Therefore the term 
\begin{eqnarray}
\left( P_{3}\right) _{A_{3}}^{A_{1}A_{2}A_{4}} &=&8\left( D_{\left(
A_{2}\right) _{0}\left( A_{3}\right) _{0}}^{00}D_{\left( A_{2}\right)
_{1}\left( A_{3}\right) _{1}}^{00}+D_{\left( A_{2}\right) _{0}\left(
A_{3}\right) _{1}}^{00}D_{\left( A_{2}\right) _{1}\left( A_{3}\right)
_{0}}^{00}\right) \\
&&-4\left( D_{\left( A_{3}\right) _{0}}^{000}-D_{\left( A_{3}\right)
_{0}}^{001}\right) \left( D_{\left( A_{3}\right) _{1}}^{000}-D_{\left(
A_{3}\right) _{1}}^{001}\right) \\
&=&2\left( I_{3}\right) _{\left( A_{3}\right)
_{0}}^{A_{1}A_{2}A_{4}}+2\left( I_{3}\right) _{\left( A_{3}\right)
_{1}}^{A_{1}A_{2}A_{4}}-2\left( I_{3}\right) _{A_{3}}^{A_{1}A_{2}A_{4}},
\end{eqnarray}%
is a three qubit invariant. Since 
\begin{equation*}
\left( I_{4}\right) _{A_{3}}^{A_{1}A_{2}A_{4}}=\left(
D^{0000}-D^{0001}+D^{0010}-D^{0011}\right) ^{2}-4\left( D_{\left(
A_{2}\right) _{0}}^{000}-D_{\left( A_{2}\right) _{0}}^{001}\right) \left(
D_{\left( A_{2}\right) _{1}}^{000}-D_{\left( A_{2}\right) _{1}}^{001}\right)
,
\end{equation*}%
is also $A_{1}A_{2}A_{4}$ invariant, $J_{4}^{\left( A_{1}A_{4}\right) }$ in
terms of $A_{1}A_{2}A_{4}$ invariants reads as%
\begin{equation*}
J_{4}^{\left( A_{1}A_{4}\right) }=\left( I_{4}\right)
_{A_{3}}^{A_{1}A_{2}A_{4}}+\left( P_{3}\right) _{A_{3}}^{A_{1}A_{2}A_{4}}.
\end{equation*}%
Alternatively, it is also the sum of two $A_{1}A_{3}A_{4}$ invariants. In
general, a four qubit invariants \ $J_{4}^{\left( A_{p}A_{q}\right) }$ can
be expressed in terms of three qubit invariants of sub-system $%
A_{p}A_{q}A_{r}$, \ or $A_{p}A_{q}A_{s}$. Three qubit invariants can be
manipulated by unitary transformation on the fourth qubit.

Four qubit invariant obtained by combining the invariants of U$^{A_{1}}$U$%
^{A_{3}}$ is 
\begin{eqnarray}
J_{4}^{\left( A_{1}A_{3}\right) } &=&\left(
D^{0000}-D^{0010}+D^{0001}-D^{0011}\right) ^{2}  \notag \\
&&+8\left( D_{\left( A_{2}\right) _{0}\left( A_{4}\right)
_{0}}^{00}D_{\left( A_{2}\right) _{1}\left( A_{4}\right)
_{1}}^{00}+D_{\left( A_{2}\right) _{1}\left( A_{4}\right)
_{0}}^{00}D_{\left( A_{2}\right) _{0}\left( A_{4}\right) _{1}}^{00}\right) 
\notag \\
&&-4\left( D_{\left( A_{2}\right) _{0}}^{000}-D_{\left( A_{2}\right)
_{0}}^{010}\right) \left( D_{\left( A_{2}\right) _{1}}^{000}-D_{\left(
A_{2}\right) _{1}}^{010}\right)  \notag \\
&&-4\left( D_{\left( A_{4}\right) _{0}}^{000}-D_{\left( A_{4}\right)
_{0}}^{001}\right) \left( D_{\left( A_{4}\right) _{1}}^{000}-D_{\left(
A_{4}\right) _{1}}^{001}\right) ,  \label{j13}
\end{eqnarray}%
and starting with U$^{A_{1}}$U$^{A_{2}}$ invariants we get%
\begin{eqnarray}
J_{4}^{\left( A_{1}A_{2}\right) } &=&\left(
D^{0000}-D^{0100}+D^{0010}-D^{0110}\right) ^{2}  \notag \\
&&+8D_{\left( A_{3}\right) _{0}\left( A_{4}\right) _{0}}^{00}D_{\left(
A_{3}\right) _{1}\left( A_{4}\right) _{1}}^{00}+8D_{\left( A_{3}\right)
_{1}\left( A_{4}\right) _{0}}^{00}D_{\left( A_{3}\right) _{0}\left(
A_{4}\right) _{1}}^{00}  \notag \\
&&-4\left( D_{\left( A_{3}\right) _{0}}^{000}-D_{\left( A_{3}\right)
_{0}}^{010}\right) \left( D_{\left( A_{3}\right) _{1}}^{000}-D_{\left(
A_{3}\right) _{1}}^{010}\right)  \notag \\
&&-4\left( D_{\left( A_{4}\right) _{0}}^{000}-D_{\left( A_{4}\right)
_{0}}^{010}\right) \left( D_{\left( A_{4}\right) _{1}}^{000}-D_{\left(
A_{4}\right) _{1}}^{010}\right) .  \label{j12}
\end{eqnarray}%
These invariants satisfy the condition 
\begin{equation}
\left( \left( T_{4}\right) ^{A_{1}A_{2}A_{3}A_{4}}\right) ^{2}=\frac{1}{3}%
\left( J_{4}^{\left( A_{1}A_{2}\right) }+J_{4}^{\left( A_{1}A_{3}\right)
}+J_{4}^{\left( A_{1}A_{4}\right) }\right) ,
\end{equation}%
and are used to define entanglement monotone \ 
\begin{equation}
\beta _{4}=\frac{1}{6}\sum\limits_{m<n}\beta _{4}^{\left( A_{m}A_{n}\right)
};\quad \beta _{4}^{\left( A_{m}A_{n}\right) }=\frac{4}{3}\left\vert
J_{4}^{\left( A_{m}A_{n}\right) }\right\vert .
\end{equation}%
Cosider the entangled states 
\begin{equation*}
\left\vert B\right\rangle =a\left\vert 0000\right\rangle +b\left\vert
1100\right\rangle +c\left\vert 0011\right\rangle +d\left\vert
1111\right\rangle ,
\end{equation*}%
characterized by $\tau _{4}=4\left\vert ad+bc\right\vert ^{2}$ , $%
J_{4}^{\left( A_{1}A_{2}\right) }=J_{4}^{\left( A_{3}A_{4}\right) }=\left(
ad+bc\right) ^{2}+8abcd$, and $J_{4}^{\left( A_{1}A_{4}\right)
}=J_{4}^{\left( A_{1}A_{3}\right) }=\left( ad-bc\right) ^{2}$. If $ad=bc$
then $\tau _{4}=16\left\vert ad\right\vert ^{2}$, but $J_{4}^{\left(
A_{1}A_{4}\right) }=J_{4}^{\left( A_{1}A_{3}\right) }=0$ and the state 
\begin{equation*}
\left\vert B\right\rangle _{ad=bc}=\left( a\left\vert 00\right\rangle
+b\left\vert 11\right\rangle \right) \left( \left\vert 00\right\rangle +%
\frac{c}{a}\left\vert 11\right\rangle \right) ,
\end{equation*}%
is a product of two qubit entangled states.

\subsection{Sextic Invariant}

Set of transformation equations for negativity fonts can be used to obtain
additional invariants to discriminate between different types of quantum
correlations in four qubit states. In this section an expression for degree
six invariant, obtained from set of transformation equations for negativity
fonts is given. A sextic invariantes $\left( I_{6}^{\left( A_{p}Aq\right)
}\right) ^{A_{p}A_{q}A_{r}A_{s}}$ may be constructed by starting with a
product of three invariants of $U^{A_{p}}U^{A_{q}}$ containing determinants
of negativity fonts in $\rho _{G}^{T_{A_{p}}}$. For instance, transformation
Eqs. (\ref{t1}-\ref{t4}), when used to construct an invariant by starting
from a product of three invariants of $U^{A_{2}}U^{A_{3}}$ containing
determinants of negativity fonts in $\rho _{G}^{T_{A_{2}}}$, yield the
invariant%
\begin{eqnarray*}
\left( I_{6}^{\left( A_{2}A_{3}\right) }\right) ^{A_{1}A_{2}A_{3}A_{4}}
&=&D_{\left( A_{1}\right) _{0}\left( A_{4}\right) _{0}}^{00}D_{\left(
A_{1}\right) _{1}\left( A_{4}\right) _{1}}^{00}\left(
D^{0000}+D^{0001}-D^{0010}-D^{0011}\right) \\
&&-D_{\left( A_{1}\right) _{0}\left( A_{4}\right) _{1}}^{00}D_{\left(
A_{1}\right) _{1}\left( A_{4}\right) _{0}}^{00}\left(
D^{0000}+D^{0001}-D^{0010}-D^{0011}\right) \\
&&+D_{\left( A_{1}\right) _{0}\left( A_{4}\right) _{1}}^{00}\left( D_{\left(
A_{1}\right) _{1}}^{000}-D_{\left( A_{1}\right) _{1}}^{100}\right) \left(
D_{\left( A_{4}\right) _{0}}^{000}-D_{\left( A_{4}\right) _{0}}^{001}\right)
\\
&&-D_{\left( A_{1}\right) _{0}\left( A_{4}\right) _{0}}^{00}\left( D_{\left(
A_{1}\right) _{1}}^{000}-D_{\left( A_{1}\right) _{1}}^{010}\right) \left(
D_{\left( A_{4}\right) _{1}}^{000}-D_{\left( A_{4}\right) _{1}}^{010}\right)
\\
&&+D_{\left( A_{1}\right) _{1}\left( A_{4}\right) _{0}}^{00}\left( D_{\left(
A_{1}\right) _{0}}^{000}-D_{\left( A_{1}\right) _{0}}^{010}\right) \left(
D_{\left( A_{4}\right) _{1}}^{000}-D_{\left( A_{4}\right) _{1}}^{010}\right)
\\
&&-D_{\left( A_{1}\right) _{1}\left( A_{4}\right) _{1}}^{00}\left( D_{\left(
A_{1}\right) _{0}}^{000}-D_{\left( A_{1}\right) _{0}}^{010}\right) \left(
D_{\left( A_{4}\right) _{0}}^{000}-D_{\left( A_{4}\right) _{0}}^{100}\right)
,
\end{eqnarray*}%
which is the same as invariant $D_{xt}$ of ref. \cite{luqu06}. However, when
expressed in terms of negativity fonts, each term gives a clear picture of
how negativity fonts may be distributed in the state to generate a non-zero$%
\left( I_{6}^{\left( A_{2}A_{3}\right) }\right) ^{A_{1}A_{2}A_{3}A_{4}}$.
Additional degree six invariants can be obtained similarly. The power of
sextic invariant lies in distinguishing between states for which degree four
invariants have the same value.

\section{Maximally entangled Four qubit states}

The maximally entangled four qubit GHZ \cite{gree89} state 
\begin{equation}
\left\vert \Psi _{GHZ}\right\rangle =\frac{1}{\sqrt{2}}\left( \left\vert
0000\right\rangle +\left\vert 1111\right\rangle \right) ,  \label{ghz}
\end{equation}%
is characterized by a single $4-$way negativity font with determinant $%
D^{0000}=a_{0000}a_{1111}=\frac{1}{2}$, which corresponds to $\tau
_{4}=1,\beta _{4}^{A_{p}A_{q}}=\frac{1}{3}$. The state has only four-way
correlations therefore $\rho _{G}^{T_{p}}=\rho _{4}^{T_{p}}$, and $\left(
N_{G}^{A_{p}}\right) ^{2}=\tau _{4}$ for ($p=1-4$). The value of degree six
invariant $I_{6}^{\left( A_{2}A_{3}\right) }=0$ for this state.

To characterize the entanglement of state 
\begin{eqnarray}
\left\vert \chi \right\rangle &=&\frac{1}{\sqrt{8}}\left( \left\vert
0000\right\rangle -\left\vert 0011\right\rangle +\left\vert
0110\right\rangle -\left\vert 0101\right\rangle \right)  \notag \\
&&+\frac{1}{\sqrt{8}}\left( \left\vert 1100\right\rangle +\left\vert
1111\right\rangle +\left\vert 1010\right\rangle +\left\vert
1001\right\rangle \right) ,
\end{eqnarray}%
expectation values of third, fourth and sixth order filter operators \cite%
{oste05,oste06} have been used in ref. \cite{yeo06} \ and the equivalence of
the state to some graph states demonstrated \cite{ye08}. We verify that the
state $\left\vert \chi \right\rangle $ is characterized by $\tau _{4}=0$, $%
J^{A_{1}A_{2}}=J^{\left( A_{1}A_{3}\right) }=J^{\left( A_{2}A_{4}\right)
}=J^{\left( A_{3}A_{4}\right) }=-\frac{1}{4}$, and $J^{\left(
A_{1}A_{4}\right) }=J^{\left( A_{2}A_{3}\right) }=\frac{1}{2}$. Therefore,
the state has $\beta _{4}^{A_{1}A_{2}}=\beta _{4}^{A_{1}A_{3}}=\beta
_{4}^{A_{2}A_{4}}=\beta _{4}^{A_{3}A_{4}}=\frac{1}{3}$, while $\beta
_{4}^{A_{1}A_{4}}=\beta _{4}^{A_{2}A_{3}}=\frac{2}{3}$, indicating that the
entanglement of state $\left\vert \chi \right\rangle $ is distinct from that
of $\left\vert \Psi _{GHZ}\right\rangle $ $\left( \tau _{4}=1,\beta
_{4}^{A_{1}A_{2}}=\beta _{4}^{A_{1}A_{3}}=\beta _{4}^{A_{1}A_{4}}=\frac{1}{3}%
\right) $. Negativity font formalism provides an easy way to determine the
local unitary transformations that transform the state $\left\vert \chi
\right\rangle $ to canonical form that is a state written in terms of
minimum number of local basis product states \cite{acin00}. In general, by
examining the determinants of negativity fonts that contribute to a given
invariant, it is possible to use transformation equations to determine local
unitaries connecting two unitary equivalent states.

We look at the invariant $J^{A_{1}A_{2}}$ for the state $\left\vert \chi
\right\rangle $. Manifestly, the state has four-way and two way fonts,
however the only nonzero contribution to this invariant is $%
J^{A_{1}A_{2}}=8D_{\left( A_{3}\right) _{0}\left( A_{4}\right)
_{0}}^{00}D_{\left( A_{3}\right) _{1}\left( A_{4}\right)
_{1}}^{00}+8D_{\left( A_{3}\right) _{1}\left( A_{4}\right)
_{0}}^{00}D_{\left( A_{3}\right) _{0}\left( A_{4}\right) _{1}}^{00}=-\frac{1%
}{4}$. Local unitary U$^{A_{3}}=\frac{1}{\sqrt{1+\left\vert x\right\vert ^{2}%
}}\left[ 
\begin{array}{cc}
1 & -x^{\ast } \\ 
x & 1%
\end{array}%
\right] $, transforms the negativity fonts such that%
\begin{equation}
\left( D_{\left( A_{3}\right) _{0}\left( A_{4}\right) _{i_{4}}}^{00}\right)
^{\prime }=\frac{1}{1+\left\vert x\right\vert ^{2}}\left( D_{\left(
A_{3}\right) _{0}\left( A_{4}\right) _{i_{4}}}^{00}+\left( x^{\ast }\right)
^{2}D_{\left( A_{3}\right) _{1}\left( A_{4}\right) _{i_{4}}}^{00}\right) ,
\label{ua31}
\end{equation}%
\begin{equation}
\left( D_{\left( A_{3}\right) _{1}\left( A_{4}\right) _{i_{4}}}^{00}\right)
^{\prime }=\frac{1}{1+\left\vert x\right\vert ^{2}}\left( D_{\left(
A_{3}\right) _{1}\left( A_{4}\right) _{i_{4}}}^{00}+x^{2}D_{\left(
A_{3}\right) _{0}\left( A_{4}\right) _{i_{4}}}^{00}\right) .  \label{ua32}
\end{equation}%
The choice 
\begin{equation*}
\left( x^{\ast }\right) ^{2}=-\frac{D_{\left( A_{3}\right) _{0}\left(
A_{4}\right) _{0}}^{00}}{D_{\left( A_{3}\right) _{1}\left( A_{4}\right)
_{0}}^{00}}=1\text{, }
\end{equation*}%
makes $\left( D_{\left( A_{3}\right) _{i_{3}}\left( A_{4}\right)
_{i_{4}}}^{00}\right) ^{\prime }=0,$ ($i_{3},i_{4}=0,1)$ and generates $3-$%
way negativity fonts. Next, unitaries U$^{A_{1}}=U^{A_{2}}=\frac{1}{\sqrt{2}}%
\left[ 
\begin{array}{cc}
1 & -1 \\ 
1 & 1%
\end{array}%
\right] $ on qubits $A_{1}$ and $A_{2}$ transform the state to canonical form%
\begin{equation*}
\left\vert \chi \right\rangle _{c}=\frac{1}{2}\left( \left\vert
0000\right\rangle -\left\vert 0111\right\rangle +\left\vert
1110\right\rangle +\left\vert 1001\right\rangle \right) ,
\end{equation*}%
with only three and two-way negativity fonts and $J^{A_{1}A_{2}}=-\frac{1}{4}
$. Obviously, no entangled pairs $A_{1}A_{2}$ or $A_{1}A_{3}$ can be
obtained from $\left\vert \chi \right\rangle $ on state reduction. Total
number of distinct negativity fonts in $\left\vert \chi \right\rangle _{c}$
is six that is four $3-$way fonts and two $2-$way fonts. An interesting
feature of $\left\vert \chi \right\rangle _{c}$ is that $4-$way three qubit
invariants are zero\ for two of the qubits on this state.

Another four qubit state, conjectured to have maximal entanglement in ref. 
\cite{higu00}, is 
\begin{eqnarray}
\left\vert HS\right\rangle &=&\frac{1}{\sqrt{6}}\left( \left\vert
0011\right\rangle +\left\vert 1100\right\rangle +\exp \left( \frac{i2\pi }{3}%
\right) \left( \left\vert 1010\right\rangle +\left\vert 0101\right\rangle
\right) \right)  \notag \\
&&+\frac{1}{\sqrt{6}}\exp \left( \frac{i4\pi }{3}\right) \left( \left\vert
1001\right\rangle +0110\right) .
\end{eqnarray}%
Two way negativity fonts $D_{\left( A_{3}\right) _{0}\left( A_{4}\right)
_{1}}^{00}=D_{\left( A_{3}\right) _{1}\left( A_{4}\right) _{0}}^{00}=\frac{1%
}{6}$, and $4-$way negativity fonts $D^{0011}=\frac{1}{6}$, $D^{0001}=\frac{1%
}{12}\left( 1-i\sqrt{3}\right) ,$ and D$^{0010}=\frac{1}{12}\left( 1+i\sqrt{3%
}\right) $) transform under the action of U$^{A_{3}}$, U$^{A_{4}}$
generating three-way negativity fonts, however, unlike the state $\left\vert
\chi \right\rangle $, this state cannot be written in a form with only $3-$%
way and $2-$way coherences. It is found that in this case three qubit
invariants $\left( P_{3}\right) _{A_{3}}^{A_{1}A_{2}A_{4}}$ as well as $%
\left( I_{4}\right) _{A_{3}}^{A_{1}A_{2}A_{4}}$ contribute to $%
J^{A_{1}A_{2}} $. Similar observations hold for other J invariants.

Recently, Gilad and Wallach \cite{gour10} have pointed out that three
cluster states \cite{brie01,raus01}%
\begin{equation}
\left\vert C_{1}\right\rangle =\frac{1}{2}\left( \left\vert
0000\right\rangle +\left\vert 1100\right\rangle +\left\vert
0011\right\rangle -\left\vert 1111\right\rangle \right)  \label{c1}
\end{equation}%
\begin{equation}
\left\vert C_{2}\right\rangle =\frac{1}{2}\left( \left\vert
0000\right\rangle +\left\vert 0110\right\rangle +\left\vert
1001\right\rangle -\left\vert 1111\right\rangle \right) ,  \label{c2}
\end{equation}%
\begin{equation}
\left\vert C_{3}\right\rangle =\frac{1}{2}\left( \left\vert
0000\right\rangle +\left\vert 1010\right\rangle +\left\vert
0101\right\rangle -\left\vert 1111\right\rangle \right) ,  \label{c3}
\end{equation}%
are the only states that maximize the Renyi $\alpha -$entropy of
entanglement for all $\alpha \geq 2$. \ The state $\left\vert
C_{1}\right\rangle $ with $\rho _{G}^{T_{A}}=\rho _{4}^{T_{A}}+\rho
_{2}^{T_{A}}-\rho $, ($\tau _{4}=0$) can be transformed by local unitaries
on qubits A$_{1}$ and A$_{2}$ to the form%
\begin{equation*}
\left\vert C_{1}\right\rangle ^{\prime }=\left\vert 0000\right\rangle
+\left\vert 1100\right\rangle +\left\vert 1011\right\rangle +\left\vert
0111\right\rangle ,
\end{equation*}%
with $\rho _{G}^{T_{A}}=\rho _{3}^{T_{A}}+\rho _{2}^{T_{A}}-\rho $. A
similar observation holds for the states $\left\vert C_{2}\right\rangle $,
and $\left\vert C_{3}\right\rangle $. Calculation of three qubit invariants
shows that the distinguishing feature of the states $\left\vert
C_{1}\right\rangle $, $\left\vert C_{2}\right\rangle $, and $\left\vert
C_{3}\right\rangle $\ is null invariant $\left( P_{3}\right)
^{A_{p}A_{q}A_{r}}$ for two of the qubits, while $\left( I_{4}\right)
^{A_{p}A_{q}A_{r}}$ is non zero.

Another candidate for maximally entangled state, found through a numerical
search in ref. \cite{brow06}, is%
\begin{eqnarray*}
\left\vert \Phi \right\rangle &=&\frac{1}{2}\left( \left\vert
0000\right\rangle +\left\vert 1101\right\rangle \right) \\
&&+\frac{1}{\sqrt{8}}\left( \left\vert 1011\right\rangle +\left\vert
0011\right\rangle +\left\vert 0110\right\rangle -\left\vert
1110\right\rangle \right) ,
\end{eqnarray*}%
This state, just like $\left\vert \chi \right\rangle _{c}$, has only three
and two way negativity fonts. Unlike $\left\vert \chi \right\rangle _{c}$,
however, $J_{4}^{A_{1}A_{3}}=J^{A_{2}A_{4}}=0$, because $\left( I_{4}\right)
^{A_{1}A_{3}A_{2}}=-\left( P_{3}\right) ^{A_{1}A_{3}A_{2}}$.

In Table \ref{table1}, the numerical values of four qubit invariants $\left(
T_{4}\right) ^{2}$, $J_{4}^{A_{1}A_{2}}=J^{A_{3}A_{4}}$, $%
J_{4}^{A_{1}A_{3}}=J^{A_{2}A_{4}}$, and $J_{4}^{A_{1}A_{4}}=J^{A_{2}A_{3}}$,
are listed for $\left\vert GHZ\right\rangle $ state, $\left\vert \chi
\right\rangle _{c}$ state, $\left\vert HS\right\rangle $ state, cluster
states $\left\vert C_{1}\right\rangle $, $\left\vert C_{2}\right\rangle $, $%
\left\vert C_{3}\right\rangle $ \ and state $\left\vert \Phi \right\rangle $%
. Four tangle, $\beta _{4}$, average global negativity, and $\Delta _{4}$
are also included therein. Degree six invariant $I_{6}^{\left(
A_{2}A_{3}\right) }$ as well as three qubit invariants $\left( I_{3}\right)
^{A_{p}A_{q}A_{r}}$ and $\left( P_{3}\right) ^{A_{p}A_{q}A_{r}}$ are
displayed in Table \ref{table2}.

The state $\left\vert \Phi \right\rangle $ is not different from $\left\vert
HS\right\rangle $ state, as far as $4-$way correlations are concerned.
However, the degree six invariant $\left( I_{6}^{\left( A_{2}A_{3}\right)
}\right) $ is zero for the state $\left\vert \Phi \right\rangle $ The values
of degree four invariants are the same for cluster states and state $%
\left\vert \chi \right\rangle _{c}$, but these are not unitary equivalent
states. The difference between $\left\vert \Phi \right\rangle $, $\left\vert
\chi \right\rangle _{c}$ and cluster states lies in the entanglement of
three qubit subsystems as is manifest in the values of three qubit
invariants in Table \ref{table2}.

\begin{table}[t]
\caption{Numeical values of fourqubit invariants for $\left\vert
GHZ\right\rangle $, state \protect\cite{gree89} , $\left\vert \protect\chi %
\right\rangle $, state \protect\cite{yeo06,ye08}, $\left\vert
HS\right\rangle $, state \protect\cite{higu00} , cluster states $\left\vert
C_{1}\right\rangle $, $\left\vert C_{2}\right\rangle $, $\left\vert
C_{3}\right\rangle $, \protect\cite{brie01,raus01,gour10} and state $%
\left\vert \Phi \right\rangle $ \protect\cite{brow06}.}\centering
\par
\begin{tabular}{||c||c||c||c||c||c||c||c||c||}
\hline\hline
$State$ & $\left( T_{4}\right) ^{2}$ & $J_{4}^{A_{1}A_{2}}$ & $%
J_{4}^{A_{1}A_{3}}$ & $J_{4}^{A_{1}A_{4}}$ & $\tau _{4}$ & $\beta _{4}=\frac{%
1}{3}\sum\limits_{j=2}^{4}\beta _{4}^{A_{1}A_{j}}$ & $\frac{1}{4}%
\sum\limits_{p=1}^{4}\left( N_{G}^{A_{p}}\right) ^{2}$ & $\Delta _{4}$ \\ 
\hline\hline
$\left\vert GHZ\right\rangle $ & $\frac{1}{4}$ & $\frac{1}{4}$ & $\frac{1}{4}
$ & $\frac{1}{4}$ & $1$ & $\frac{1}{3}$ & $1$ & $0$ \\ \hline\hline
$\left\vert \chi \right\rangle $ & $0$ & $-\frac{1}{4}$ & $-\frac{1}{4}$ & $%
\frac{1}{2}$ & $0$ & $\frac{4}{9}$ & $1$ & $1$ \\ \hline\hline
$\left\vert HS\right\rangle $ & $0$ & $\frac{1}{3}$ & $\frac{i\sqrt{3}\ -1}{6%
}\ $ & $-\frac{i\sqrt{3}\ +1}{6}$ & $0$ & $\frac{4}{9}$ & $1$ & $1$ \\ 
\hline\hline
$\left\vert C_{1}\right\rangle $ & $0$ & $-\frac{1}{2}$ & $\frac{1}{4}$ & $%
\frac{1}{4}$ & $0$ & $\frac{4}{9}$ & $1$ & $1$ \\ \hline\hline
$\left\vert C_{2}\right\rangle $ & $0$ & $\frac{1}{4}$ & $\frac{1}{4}$ & $-%
\frac{1}{2}$ & $0$ & $\frac{4}{9}$ & $1$ & $1$ \\ \hline\hline
$\left\vert C_{3}\right\rangle $ & $0$ & $\frac{1}{4}$ & $-\frac{1}{2}$ & $%
\frac{1}{4}$ & $0$ & $\frac{4}{9}$ & $1$ & $1$ \\ \hline\hline
$\left\vert \Phi \right\rangle $ & $0$ & $\frac{3}{8}$ & $0$ & $-\frac{3}{8}$
& $0$ & $\frac{1}{3}$ & $1$ & $1$ \\ \hline\hline
\end{tabular}%
\label{table1}
\end{table}

\begin{table}[t]
\caption{Numeical values of $\left( T_{4}\right) ^{2}$ , sextic invariant $%
I_{6}^{A_{2}A_{3}},$ and three qubit invariants for $\left\vert
GHZ\right\rangle $, $\left\vert \protect\chi \right\rangle _{c}$, $%
\left\vert HS\right\rangle $ , $\left\vert C_{1}\right\rangle $, $\left\vert
C_{2}\right\rangle $, $\left\vert C_{3}\right\rangle $, and $\left\vert \Phi
\right\rangle $, States.}\centering
\par
\begin{tabular}{||c||c||c||c||c||c||c||c||c||}
\hline\hline
$State$ & $\left( T_{4}\right) ^{2}$ & $I_{6}^{A_{2}A_{3}}$ & $\left(
I_{4}\right) ^{A_{1}A_{2}A_{3}}$ & $\left( P_{3}\right) ^{A_{1}A_{2}A_{3}}$
& $\left( I_{4}\right) ^{A_{1}A_{3}A_{2}}$ & $\left( P_{3}\right)
^{A_{1}A_{3}A_{2}}$ & $\left( I_{4}\right) ^{A_{1}A_{4}A_{2}}$ & $\left(
P_{3}\right) ^{A_{1}A_{4}A_{2}}$ \\ \hline\hline
$\left\vert GHZ\right\rangle $ & $\frac{1}{4}$ & $0$ & $\frac{1}{4}$ & $0$ & 
$\frac{1}{4}$ & $0$ & $\frac{1}{4}$ & $0$ \\ \hline\hline
$\left\vert HS\right\rangle $ & $0$ & $\frac{i\sqrt{3}\ -1}{6}$ & $\frac{1}{9%
}$ & $\frac{2}{9}$ & $\frac{i\sqrt{3}\ -1}{18}$ & $\frac{i\sqrt{3}\ -1}{9}$
& $-\frac{i\sqrt{3}\ +1}{18}$ & $-\frac{i\sqrt{3}\ +1}{9}$ \\ \hline\hline
$\left\vert \Phi \right\rangle $ & $0$ & $0$ & $\frac{1}{4}$ & $\frac{1}{8}$
& $\frac{1}{8}$ & $-\frac{1}{8}$ & $-\frac{1}{8}$ & $-\frac{1}{4}$ \\ 
\hline\hline
$\left\vert \chi \right\rangle _{c}$ & $0$ & $0$ & $0$ & $-\frac{1}{4}$ & $0$
& $-\frac{1}{4}$ & $0$ & $\frac{1}{2}$ \\ \hline\hline
$\left\vert C_{1}\right\rangle $ & $0$ & $0$ & $0$ & $-\frac{1}{2}$ & $\frac{%
1}{4}$ & $0$ & $\frac{1}{4}$ & $0$ \\ \hline\hline
$\left\vert C_{2}\right\rangle $ & $0$ & $0$ & $\frac{1}{4}$ & $0$ & $\frac{1%
}{4}$ & $0$ & $0$ & $-\frac{1}{2}$ \\ \hline\hline
$\left\vert C_{3}\right\rangle $ & $0$ & $0$ & $\frac{1}{4}$ & $0$ & $0$ & $-%
\frac{1}{2}$ & $\frac{1}{4}$ & $0$ \\ \hline\hline
\end{tabular}%
\label{table2}
\end{table}

We notice that $\left\vert GHZ\right\rangle $ state, $\left\vert
HS\right\rangle $ state, $\left\vert \chi \right\rangle $ state$,$ group of
states$\ \left\vert C_{1}\right\rangle $, $\left\vert C_{2}\right\rangle $, $%
\left\vert C_{3}\right\rangle $ and the state $\left\vert \Phi \right\rangle 
$ belong to five distinct four qubit entanglement classes. Each state is
maximally entangled in its own class with $\frac{1}{4}\sum\limits_{p=1}^{4}%
\left( N_{G}^{A_{p}}\right) ^{2}=1$ for each qubit, however with different
capability for performing information processing tasks.

\section{Conclusions}

To summarize, the transformation equations for negativity fonts under
unitary transformations yield relevant $N-$qubit invariants and determine
local unitaries relating unitary equivalent states. An expression for global
negativity in terms of determinants of negativity fonts has been found. The
squared negativity of $N-$qubit partially transposed operator is four times
the sum of squared moduli of determinants of all possible negativity fonts.
The structure of four qubit invariants of degree four that detect
entanglement between pairs of qubits indicates why some of the unitary
equivalent states may have\ different sets of $K-$way coherences. It is
shown that a four qubit invariant $J_{4}^{A_{p}A_{q}}$ can be expressed in
terms of three qubit invariants for qubits $A_{p}A_{q}A_{r}$, \ or $%
A_{p}A_{q}A_{s}$. Three qubit invariants can be manipulated by unitary
transformation on the fourth qubit but their value for the canonical state
is unique. In the context of four qubit states studied in the article, the
two types of three qubit entanglement invariants, each corresponding to a
different type of quantum correlations present in the canonical state, play
an important role in distinguishing between states with inequivalent
entanlement types. Degree six invariants can also be constructed easily from
Eqs. (\ref{t1}-\ref{t4}), as shown by writing the invariant $\left(
I_{6}^{\left( A_{2}A_{3}\right) }\right) ^{A_{1}A_{2}A_{3}A_{4}}$.
Decomposition of partially transposed matrix in to $K-$way partial
transposes is a tool to identify the type of quantum correlations which
entangle the qubits. We have used the expressions of polynomial invariants
in terms of negativity fonts to elucidate the difference in microstructure
of some well known four qubit pure states. We conclude that the entanglement
in four qubit $\left\vert GHZ\right\rangle $ state, $\left\vert \chi
\right\rangle $ state, $\left\vert HS\right\rangle $ state, cluster states $%
\left\vert C_{1}\right\rangle $, $\left\vert C_{2}\right\rangle $, $%
\left\vert C_{3}\right\rangle $, and state $\left\vert \Phi \right\rangle $
is qualitatively different since the states belong to different classes of
four qubit entangled states. Cluster states $\left\vert C_{1}\right\rangle $%
, $\left\vert C_{2}\right\rangle $, $\left\vert C_{3}\right\rangle ,$ differ
from the $\left\vert \chi \right\rangle _{c}$ state,\ in having different
type of three qubit correlations in canonical form. These results indicate
that along with composite system invariants, one needs subsystem invariants
in canonical form to characterize the entanglement of a state. The four
qubit entangled states investigated here do not represent all four qubit
entanglement types represented by nine families of four qubit states \cite%
{vers02}. However, the results provide insight to formulate efficient
criterion for classification of four qubit entangled states. In ref. \cite%
{shar102}, the general method for writing N-tangle was given. In general,
for n-even square of degree two invariant, having only N-way fonts, can be
written as a sum of invariants that detect the entanglement of parts of the
composite system. As such, the ideas developed for four qubits may be
extended to multi qubit systems.

\begin{acknowledgments}
This work is supported by Faep Uel, Funda\c{c}\~{a}o Arauc\'{a}ria and CNPq,
Brazil.
\end{acknowledgments}

\end{document}